\documentstyle[12pt]{article}

\def\GeV{\,{\rm GeV}}

\def\sec{\,{\rm sec}}

\def\rcm{\,{\rm cm}}

\def\Mpc{\,{\rm Mpc}}

\def\eV{{\,\rm eV}}

\def\cmm2{{\,\rm cm^{-2}}}
\def\cm2{{\,{\rm cm}^2}}
\def\cmm3{{\,{\rm cm}^{-3}}}
\def\gcmm3{{\,{\rm g\,cm^{-3}}}}
\def\kms{\,{\rm km\,s^{-1}}}

\def\mpl{{m_{\rm Pl}}}

\def\trh{T_{\rm RH}}

\def\zl{z_{\rm LSS}}
\def\zeq{z_{\rm EQ}}

\def\la{\mathrel{\mathpalette\fun <}}
\def\ga{\mathrel{\mathpalette\fun >}}
\def\fun#1#2{\lower3.6pt\vbox{\baselineskip0pt\lineskip.9pt
  \ialign{$\mathsurround=0pt#1\hfil##\hfil$\crcr#2\crcr\sim\crcr}}}
\begin{document}
\pagestyle{empty}
\begin{center}
% \leftline{\large DRAFT}
\rightline{FERMILAB--Pub--93/069-A}
\rightline{CfPA--93}
% \rightline{astro-ph/93060**}
% \rightline{submitted to {\it Physical Review D}}

\vspace{.2in}
{\Large \bf Tensor Perturbations in Inflationary Models\\
as a Probe of Cosmology} \\

\vspace{.2in}
Michael S. Turner,$^{1,2}$ Martin White,$^3$ and James E. Lidsey$^2$\\

\vspace{.2in}

$^1${\it Departments of Physics and of Astronomy \& Astrophysics\\
Enrico Fermi Institute, The University of Chicago, Chicago, IL~~60637-1433}\\

\vspace{.1in}

$^2${\it NASA/Fermilab Astrophysics Center,
Fermi National Accelerator Laboratory, Batavia, IL~~60510-0500}\\

\vspace{.1in}

$^3${\it Center for Particle Astrophysics, University of California\\
Berkeley, CA~~94720}

\end{center}

\vspace{.3in}

\centerline{\bf ABSTRACT}
\noindent  In principle, the tensor metric (gravity-wave)
perturbations that arise
in inflationary models can, beyond probing
the underlying inflationary model, provide
information about the Universe:  ionization
history, presence of a cosmological constant,
and epoch of matter-radiation equality.  Because
tensor perturbations give rise to anisotropy of the
cosmic background radiation (CBR) solely through
the Sachs-Wolfe effect we are able to calculate
analytically the variance of
the multipole moments of this part of the CBR
temperature anisotropy.  In so doing, we carefully take
account of the contribution of tensor perturbations that
entered the Hubble radius during both the matter-dominated
and radiation-dominated epoch by means of a
transfer function.  (Previously, only
those modes that entered during the matter era were properly
taken into account.)   The striking feature in the spectrum of multipole
amplitudes is a dramatic fall off for
$l\ga \sqrt{1+\zl}$, where $\zl$ is the red shift of the
last-scattering surface, which depends upon the ionization
history of the Universe.  Finally, using our transfer function
we provide a more precise formula for the energy
density in stochastic gravitational waves from inflation, and, using
the Cosmic Background Explorer Differential
Microwave Radiometer (COBE DMR) quadrupole normalization, we express the
this energy density in terms
of the ``tilt'' of the spectrum of tensor perturbations alone and
show that it is unlikely that the stochastic
background of gravity waves can be detected directly in the foreseeable future.

\newpage
\pagestyle{plain}
\setcounter{page}{1}
\newpage
\section{Introduction}

Quantum fluctuations arising during inflation lead to
a spectrum of scalar (density) \cite{scalar}
and tensor (gravity-wave) \cite{tensor}
metric perturbations which are nearly scale invariant \cite{inflation}.
In turn, both give rise to temperature anisotropy in the
cosmic background radiation (CBR),
with their relative contributions depending upon the
steepness of the inflationary potential \cite{tilt,turner}.
CBR anisotropy and other astrophysical
data, e.g., red-shift surveys, peculiar-velocity measurements,
and data from gravity-wave detectors, can, in principle, be used
to learn much about the inflationary potential in the
narrow interval that governs the modes that affect
astrophysically interesting scales.  For example,
they can be used to infer the value
of the inflationary potential, its steepness and the change in
its steepness \cite{turner,reconstruct}.

In all likelihood CBR anisotropy provides the cleanest
and most sensitive probe of
metric perturbations.  The CBR anisotropies that
arise due to scalar and gravity-wave fluctuations
add incoherently and can thus be computed independently.
The calculation of the anisotropy that arises due
to scalar perturbations is both complicated and well
understood:  anisotropy
arises due to at least three physical effects,
the Sachs-Wolfe effect \cite{sw}
(gravitational potential differences on the last-scattering
surface), the velocity of the last-scattering surface,
and intrinsic fluctuations in the CBR temperature at last scattering.
Further, the ionization history and
baryon density are very important \cite{cbranisotropy}.
(On large-angular scales, for standard recombination
$\theta\ga 2^\circ$, the Sachs-Wolfe effect dominates,
and on small-angular scales the other two effects dominate.)

The CBR anisotropy
due to tensor perturbations arises solely from the Sachs-Wolfe
effect.  It depends significantly upon the
red shift of the last-scattering surface, and less significantly
upon a possible
cosmological constant and the red shift of matter-radiation
equality (through the value of the Hubble constant).
Thus, {\it if} the tensor contribution to CBR
anisotropy can be separated from the scalar
contribution, it provides a very direct probe of cosmology.
The tensor contribution has been computed analytically, but only
on large-angular scales, with other simplifying
assumptions being made \cite{cbrtensor}, and recently, it
has been computed numerically in the case
of standard recombination and a matter-dominated Universe \cite{bondetal}.

The purpose of our paper to give simple and accurate analytic
formulae for the tensor contribution to the variance in
the CBR temperature multipoles
$\langle |a_{lm}|^2\rangle$ (``angular power spectrum'').
In the previous analytical work \cite{cbrtensor}, only the modes that
enter the horizon after matter-radiation
equality were properly taken into account, so that these
results are accurate only on large angular scales;
we take into account the modes that enter the
horizon before matter-radiation
equality by means of a transfer function and
thereby accurately describe the CBR anisotropy
that arises on all angular scales.  The transfer function also allows us to
give a more precise expression for the energy density
in stochastic gravity waves produced by inflation, and, unfortunately,
we show that it is very unlikely that this background
can be detected directly.
We compare our results for CBR anisotropy, where possible, to the
numerical results of Ref.~\cite{bondetal}, and
discuss how the tensor multipoles depend upon
the underlying cosmological parameters:
the red shift of last scattering, $\zl$; the
red shift of matter-radiation equality, $\zeq$;
and the presence of a cosmological constant.

\section{Gravity Waves and CBR Anisotropy}

\subsection{Qualitative view}

In this Section we begin simply and gradually add detail,
ending with our most general formulae.  In that spirit we
will assume scale-invariant metric perturbations to begin,
and then generalize our results to allow for deviation
from scale invariance.  For scale-invariant
perturbations the amplitude of density perturbations
and gravity-wave perturbations are independent of scale
at horizon crossing in the post-inflationary
Universe:  $(\delta\rho /\rho )_{\rm HOR}\equiv\varepsilon_S$ and
$h_{\rm GW}=\varepsilon_T$.  Horizon crossing is defined
by $k_{\rm phys} =k/R \simeq H$; $R$ is the
cosmic-scale factor and $H$ is the expansion rate.  Throughout
we take the scale factor to be unity at the present epoch, so that
comoving wavenumber $k$ and physical wavenumber $k_{\rm phys}$
are equal today; comoving wavelength $\lambda$ and wavenumber
are related by:  $\lambda =2\pi /k$.

It is useful to define the conformal time today,
$\tau_0$, and at the last-scattering event,
$\tau_{\rm LSS}$:
\begin{eqnarray}
\tau_0 & = & \int_0^{t_0}\,dt/R(t) = 2H_0^{-1} ; \\
\tau_{\rm LSS} & = & \int_0^{t_{\rm LSS}}\,dt/R(t)
 \simeq \tau_0/\sqrt{1+\zl};
\end{eqnarray}
where we assume a flat, matter-dominated Universe today.
The quantities $\tau_0 \simeq
6000h^{-1}\Mpc$ and $\tau_{\rm LSS}\simeq 6000h^{-1}\Mpc /\sqrt{1+\zl}$
correspond to the comoving size
of the present horizon and that at last scattering.
The comoving distance to the last-scattering surface
$d_{\rm LSS} = \tau_0(1-1/\sqrt{1+\zl}) \simeq \tau_0$,
and thus the angle subtended by the horizon scale at
last scattering corresponds to $\theta_{\rm LSS}
\sim \tau_{\rm LSS}/\tau_0$ ($\sim 2^\circ$ for
$\zl =1100$).

In the standard picture recombination occurs at a red shift of order
1300, i.e., the ionization fraction $X_e$ becomes small,
and last-scattering occurs
a red shift of about $\zl \simeq 1100$, i.e., the photon
mean-free path becomes greater than the Hubble scale \cite{joneswyse}.
If the Universe remains
ionized much later, or is reionized after recombination,
the last-scattering event can occur much later,
$1+\zl \simeq (0.03X_e \Omega_B h)^{-2/3}$, where $\Omega_B$ is
the fraction of critical density contributed by
baryons, $H_0=100h \kms\sec^{-1}\Mpc^{-1}$, and $\Omega_0=1$.
Taking $h=0.4$, $X_e=1$, and saturating the primordial
nucleosynthesis bound to the baryon mass density,
$\Omega_Bh^2\la 0.02$ \cite{walkeretal},
last scattering could occur as late as $\zl\simeq 76$.
While the conventional wisdom has it that cold dark matter
models lack a plausible mechanism for reionizing, or keeping
the Universe ionized at $z\la 1300$, it has been suggested
that a very early generation of stars could have reionized
the Universe at red shift of order 100 or so \cite{silk}.

The physics underlying the Sachs-Wolfe effect
is very simple:  The temperature fluctuation on a given
angular scale is roughly equal to the metric fluctuation
on the corresponding length scale on the last-scattering surface.
For tensor perturbations, the metric perturbation is equal to
the gravity-wave amplitude.
For scalar perturbations the metric perturbation is given
by the fluctuation in the Newtonian potential; on
length scales that have yet
to enter the horizon at last scattering, or that entered
the horizon after matter-radiation equality,
the fluctuation in the Newtonian potential is given by
the horizon-crossing amplitude of the density perturbation.
For scales that enter the horizon while the Universe is
still radiation dominated the fluctuation in the Newtonian
potential decreases after horizon crossing, as $R^{-1}$.

The CBR temperature fluctuation that arises
on a given angular scale due to scalar perturbations
through the Sachs-Wolfe effect only is roughly:
\begin{eqnarray}\label{eq:SW}
\left({\delta T\over T}\right)_\theta & \sim &
\left({\delta\rho\over \rho}\right)_{{\rm HOR},\,k(\theta )}\qquad\ \ \ \
\simeq \qquad\qquad \varepsilon_S \qquad
\theta\ga \theta_{\rm EQ} ; \nonumber \\
\left({\delta T\over T}\right)_\theta & \sim &
\left({k_{\rm EQ}\over k}\right)\,
\left({\delta\rho\over
\rho}\right)_{{\rm HOR},\,k(\theta )}
\simeq\ \  \left({\theta\over \theta_{\rm EQ}}\right)\varepsilon_S
\qquad \theta\la \theta_{\rm EQ} ;
\end{eqnarray}
where $k(\theta )$ corresponds to the wavelength
that subtends an angle $\theta$ on the last-scattering surface,
\begin{equation}
k(\theta )  \sim \left({200^\circ \over \theta}\right)\, \tau_0^{-1} .
\end{equation}
Here, $k_{\rm EQ}$ is the scale that crossed
the horizon at matter-radiation equality, and
$\theta_{\rm EQ} \sim 1/\sqrt{1+z_{\rm EQ}}\sim 0.3^\circ$.

For scale-invariant perturbations and $\theta\ga
\theta_{\rm EQ}\sim 0.3^\circ$, the Sachs-Wolfe
contribution to the temperature fluctuation is independent
of angular scale.  On small-angular scales, $\theta\la
\theta_{\rm EQ}$, the Sachs-Wolfe contribution to
the CBR temperature anisotropy decreases, but
is a subdominant contribution to total anisotropy
produced by scalar perturbations.  On angular scales larger than
the horizon at last scattering, $\theta\ga\theta_{\rm LSS}\sim
1/\sqrt{1+z_{\rm LSS}}$ ($\sim 2^\circ$ for $\zl =1100$),
the Sachs-Wolfe effect is the dominant contribution to CBR anisotropy.

The CBR temperature fluctuation on a given angular scale
due to tensor perturbations arises through
the Sachs-Wolfe effect and is roughly:
\begin{equation}
\left({\delta T\over T}\right)_\theta  \sim
h_{\rm GW}[\zl ,k(\theta )];
\end{equation}
where $h_{\rm GW}[\zl ,k(\theta )]$ is the (dimensionless)
gravity-wave amplitude on the scale $k(\theta )$
at last scattering.  For gravity-wave modes
that have yet to re-enter the horizon at last scattering,
$h_{\rm GW}(\zl ,k)$ is just equal to $\varepsilon_T$;
once a mode enters the horizon its amplitude red shifts
as the scale factor, so that $h_{\rm GW}(\zl ,k) =
\varepsilon_T (R_{\rm HOR}/R_{\rm LSS})$.  For modes
that enter the horizon during the matter-dominated epoch,
$R_{\rm HOR}/R_{\rm LSS} \simeq (k\tau_{\rm LSS})^{-2}$; for
modes that enter during the radiation-dominated epoch,
$R_{\rm HOR}/R_{\rm LSS} \simeq (k/k_{\rm EQ})(k\tau_{\rm LSS})^{-2}$.
[A simple way to understand why the amplitude of a gravity-wave mode
red shifts as the scale factor after it enters the horizon
is that gravity-wave perturbations
correspond to massless bosons (gravitons)
whose energy density ($\propto \mpl^2k_{\rm phys}^2h_{\rm GW}^2$)
red shifts as $R^{-4}$.]

For scale-invariant tensor perturbations the CBR anisotropy
due to gravity waves is only independent of scale for
$\theta \ga \theta_{\rm LSS}$; on smaller angular scales
it decreases,
\begin{eqnarray}
\left({\delta T\over T}\right)_\theta & \sim &
\varepsilon_T \qquad\qquad\qquad\  \theta \ga \theta_{\rm LSS};\nonumber \\
\left({\delta T\over T}\right)_\theta & \sim &
\varepsilon_T\,\left({\theta \over \theta_{\rm LSS} }\right)^2
\qquad \theta_{\rm LSS} \ga \theta \ga \theta_{\rm EQ};\nonumber \\
\left({\delta T\over T}\right)_\theta & \sim &
\varepsilon_T\,\left( \theta \theta_{\rm EQ} \over
\theta_{\rm LSS}^2 \right) \qquad  \theta \la \theta_{\rm EQ} .
\end{eqnarray}

On angular scales greater than that subtended by the
horizon at last scattering, $\theta_{\rm LSS}\sim
2^\circ$ in the standard
scenario, the ratio of the tensor to the scalar contributions
to the temperature anisotropy is constant and equal
to $\varepsilon_T/\varepsilon_S$; in turn, this ratio is
related to the steepness of the inflationary potential,
evaluated when these scales crossed outside the
horizon during inflation (about 50 e-folds before
the end of inflation):
$\varepsilon_T/\varepsilon_S \sim x_{50} \equiv
(\mpl V^\prime /V)_{50}$ \cite{tilt}.  On smaller angular
scales the tensor contribution falls as $\theta^2$ because
these scales are dominated by gravity-wave modes that
have entered the horizon before the epoch of last
scattering and have had their amplitudes red shifted.
On the smallest angular scales the tensor contribution
only decreases as $\theta$, as these scales are dominated
by gravity-wave modes that enter the horizon before matter
radiation equality.  The steep fall off of gravity-wave
modes for $l\ga \sqrt{1+\zl}$ provides a potential
signature of $\zl$ and tensor perturbations.

It is conventional to expand the
CBR temperature fluctuation on the sky in spherical harmonics:
\begin{equation}
{\delta T({\bf{\hat x}}) \over T_0} = \sum_{l = 2}^{\infty}
\sum_{m=-l}^{l} a_{lm}({\bf r}) Y_{lm} ({\bf {\hat x}});
\end{equation}
where the unobservable monopole term and the dipole term, which is
dominated by the contribution of the observer's
peculiar velocity, are omitted.  The multipole
amplitudes depend upon the observer's position ${\bf r}$.
The quantity $|a_{lm}|^2$ averaged over all observation positions
(the ensemble average) is referred to as the angular
power spectrum,\footnote{More precisely, the average
is over all realizations
of the fluctuation spectrum; we have implicitly assumed spatial
ergodicity.} and is related approximately to the CBR temperature
fluctuation by
\begin{equation}
\left( {\delta T\over T} \right)^2_\theta \sim l^2\langle
|a_{lm}|^2\rangle
\ \ \ \ {\rm for}\ l\sim 200^\circ /\theta .
\end{equation}

To be more precise, the {\it rms} temperature
fluctuation averaged over the sky for a given experiment
is given by
\begin{equation}
\Biggl\langle \left({\delta T\over T_0 }
\right)^2\Biggr\rangle = \sum_{l\ge 2}
{2l+1\over 4\pi}\langle |a_{lm}|^2\rangle F_l;
\end{equation}
where $F_l$ is the appropriate response function for the
experiment.  For a two-beam experiment, where
the temperature difference between two antennas of gaussian
beam width $\sigma$ separated by angle $\theta$ is measured,
$$F_l = 2\left[ 1-P_l(\cos\theta )\right]e^{-(l+1/2)^2\sigma^2}.$$

For scale-invariant density perturbations the Sachs-Wolfe
contribution to $l^2$ times the angular
power spectrum is roughly constant and equal to $\varepsilon_S$
for $l\la l_{\rm EQ}$; thereafter it decreases as
$l^{-2}$.  For scale-invariant gravity-wave perturbations
$l^2$ times the angular power spectrum is constant
for $l\la l_{\rm LSS}$ ($\sim 30$ for standard recombination);
it decreases as $l^{-4}$ for $l\la l_{\rm EQ}$; and
and as $l^{-2}$ for $l\ga l_{\rm EQ}$.

Finally, if the scalar and tensor
perturbations are ``tilted''---that is
not scale invariant---say $\varepsilon_S \propto \lambda^{\alpha_S}$
and $\varepsilon_T \propto \lambda^{\alpha_T}$, the previous results
for $(\delta T/T)_\theta$ are modified by factors of $\theta^{\alpha_S}$ and
$\theta^{\alpha_T}$ respectively, and for
$\langle |a_{lm}|^2\rangle$, by factors of
$l^{-2\alpha_S}$ and $l^{-2\alpha_T}$ respectively.
The quantities $\alpha_S$
and $\alpha_T$ are related to the power-law indices often
used to characterize the scalar and tensor perturbations
(see below) by
$$\alpha_S = (1-n)/2; \qquad \alpha_T = -n_T/2.$$
The qualitative behaviour
of the CBR anisotropy due to scalar and tensor perturbations
is shown in Fig.~1.

\subsection{Quantitative view}

Some of what follows is a quick review of previous treatments
included for completeness; for more details
see Refs.~\cite{cbrtensor}.
To begin, we write the line element for a flat,
Friedmann-Robertson-Walker (FRW) cosmology in conformal
form plus a small perturbation $h_{\mu\nu}$:
\begin{equation}
g_{\mu\nu} = R^2(\tau ) [\eta_{\mu\nu} + h_{\mu\nu}],
\end{equation}
where $\eta_{\mu\nu}= {\rm diag}(1,-1,-1,-1)$ and
$\tau$ is conformal time.  Here we are
only interested in gravity-wave perturbations and
work in the transverse-traceless gauge, where the two independent
polarization states are $\times$ and $+$, and
$h_{00}=h_{0j}=0$.

It is simple to solve for the evolution of the cosmic-scale
factor in terms of conformal time:
\begin{equation}
R(\tau ) = \left[ \tau /\tau_0 + R_{\rm EQ}^{1/2} \right]^2 - R_{\rm EQ};
\end{equation}
where we have assumed a flat Universe with matter and radiation
(i.e., photons with present temperature 2.73K and three massless
neutrino species), $R_0 =1$, $\tau_0 =2H_0^{-1}\sqrt{1+R_{\rm EQ}}
\simeq 2H_0^{-1}$, $R_{\rm EQ}
= 4.18\times10^{-5}\,h^{-2}$ is the value of the cosmic
scale factor at the epoch of matter-radiation equality,
and $\tau_{\rm EQ} = [\sqrt{2}-1]R_{\rm EQ}^{1/2}\tau_0$.
(The conformal age of the Universe today differs from
$2H_0^{-1}$ by a small amount due to the tiny contribution
of the radiation energy density today:
$$
\tau_{\rm today} = \tau_0\left[\sqrt{1+R_{\rm EQ}}-\sqrt{R_{\rm EQ}}
\right] \simeq 2H_0^{-1}\left( 1-\sqrt{R_{\rm EQ}}\right)
\approx 2H_0^{-1}\left[ 1- {\cal O}(1\%)\right] ,
$$
which we shall henceforth neglect.)

We expand the gravity-wave perturbation in plane waves
\begin{equation}
h_{jk}({\bf x},\tau ) = (2\pi )^{-3}\int \,
d^3{k}\ h_{\bf k}^i(\tau ) \epsilon^i_{jk}\,e^{-i{\bf k}
\cdot {\bf x}} ;
\end{equation}
where $\epsilon^i_{jk}$ is the polarization tensor
and $i=\times$, $+$.  The gravity-wave perturbation
satisfies the massless Klein-Gordon equation
\begin{equation}\label{eq:kg}
{\ddot h}_{\bf k}^i +2\left({\dot R\over R}\right)
{\dot h}_{\bf k}^i +k^2h_{\bf k}^i =0;
\end{equation}
where an overdot indicates a derivative with respect to conformal time
and $k^2={\bf k}\cdot {\bf k}$.

The growing-mode solutions to this equation have simple qualitative
behaviour:  before horizon crossing ($k\tau \ll 1$)
$h_{\bf k}^i(\tau )$ is constant; well after horizon
crossing ($k\tau\gg 1$) $h_{\bf k}^i \propto \cos k\tau /R$.  For modes that
cross inside the horizon during the radiation-dominated
era the exact solution prior to and including the radiation-dominated
era is $j_0(k\tau )$; for modes that cross inside
the horizon during the matter-dominated
era the exact solution is $3j_1(k\tau )/k\tau$.  Here
$j_0(z) =\sin z/z$ and $j_1(z) = \sin z/z^2 - \cos z/z$
are the spherical-Bessel functions of
order zero and one respectively, and both Bessel-function
solutions have been normalized to unity for
$\tau \rightarrow 0$.

Well into the matter-dominated era the temporal behaviour
of modes that entered the horizon during the radiation-dominated
era is also given by $3j_1(k\tau )/k\tau$.
Thus, for $\tau\gg \tau_{\rm EQ}$ the temporal behaviour of
{\it all} modes is given by $3j_1(k\tau )/k\tau$, and it is useful to write
\begin{equation}
h_{\bf k}^i (\tau ) = h_{\bf k}^i (0)\, T(k/k_{\rm EQ})\,
\left( {3j_1(k\tau ) \over k\tau}\right) ;
\end{equation}
where the ``transfer function'' for gravitational waves,
$T(k/k_{\rm EQ})$, is only a function of $k/k_{\rm EQ}$, and
\begin{equation}
k_{\rm EQ} \equiv \tau_{\rm EQ}^{-1} ={\tau_0^{-1}R_{\rm EQ}^{-1}
\over (\sqrt{2} -1) } \simeq 6.22\times 10^{-2}h^2\Mpc^{-1};
\end{equation}
is the scale that entered the horizon at matter-radiation
equality.  Note, during the oscillatory phase ($k\tau \gg1$),
$3j_1(k\tau )/k\tau \rightarrow 3\cos k\tau /(k\tau )^2$;
in defining and computing the transfer function
we have neglected the phase of the graviton oscillations, which,
for our purposes, is not important.

The transfer function has been calculated by integrating
Eq. (\ref{eq:kg}) numerically from $\tau =0$ to $\tau =\tau_0$;
a good fit to the transfer function is
\begin{equation}
T(y) = [1.0 + 1.34y + 2.50y^2]^{1/2};
\end{equation}
where $y = k/k_{\rm EQ}$.  It might seem that one could
have computed the transfer function at any time after the
Universe becomes matter dominated and obtained the same
result.  However, as we shall emphasize again later,
the Universe becomes matter dominated more slowly than one
might have expected, and for this reason the transfer function calculated
at an earlier epoch is different.  Only well into the
matter-dominated epoch, when the radiation content is
very negligible, does it take this functional form.
Using the fact that once a mode is well inside the horizon
$h_{\bf k}^i(\tau ) \propto \cos k\tau /R$ it follows that
for modes with $k\gg k_{\rm EQ}$ the
transfer function at an earlier epoch is related to that today by
\begin{equation}
T_\tau (k/k_{\rm EQ}) = (\tau /\tau_0)^2R^{-1} T(k/k_{\rm EQ})
= \left[ 1 + 2x -2\sqrt{x+x^2} \right]T_0(k/k_{\rm EQ});
\end{equation}
where $x=R_{\rm EQ}/R(\tau )$, $T_\tau$ is the transfer
function at conformal time $\tau$, and $T$ is the transfer
function today.  The evolution of the transfer function is shown in Fig.~2.

Once a mode has crossed inside the horizon one can sensibly
talk about the energy density in gravitons associated
with that mode; it is given by
\begin{equation}\label{eq:gwdensity}
k{d\rho_i \over dk} = {\mpl^2 k^5 \over 32\pi^3 R^2}\,
\overline{|h_k^i(\tau )|^2} ;
\end{equation}
where $\overline{|h_k^i|^2}$ is the average of $|h_{\bf k}^i|^2$
over all directions $\bf{\hat k}$ and over several periods.

Inflation-produced tensor perturbations are
stochastic in nature and characterized by
a gaussian random variable.  The Fourier components
$h_{\bf k}^i $ are drawn from a distribution
whose statistical expectation is
\begin{equation}\label{eq:gwps}
\langle h_{\bf k}^i h_{\bf q}^j \rangle =
P_T (k) (2\pi )^6 \delta^{(3)} ({\bf k} -{\bf q})\delta_{ij};
\end{equation}
where the gravity-wave power spectrum is defined as
\begin{eqnarray}
P_T(k) & = & A_T k^{-3} |T(k)|^2 \left[ {3j_1(k\tau )\over k\tau}
\right]^2 ; \nonumber \\
A_T & = & {8 \over 3\pi }\, {V_{50} \over \mpl^4}\, .
\end{eqnarray}
The quantity $V_{50}$ is the value of the scalar potential
around 50 e-folds before the end of inflation when
the modes of astrophysical interest, say $\lambda\sim
1\Mpc - 10^4\Mpc$, were excited and crossed outside the
horizon; see \cite{turner}.  [We note that
the ensemble average and power spectrum are related by,
$\langle |h_{\bf k}^i|^2 \rangle = (2\pi )^3P_T(k)$.]

Using Eqs.~(\ref{eq:gwdensity}, \ref{eq:gwps}) it is a
simple matter to compute the
energy density today in stochastic gravitational-wave
background produced by inflation:
\begin{equation}
{d\Omega_{\rm GW} \over d\ln k} \equiv
\sum_i {k\over \rho_{\rm crit}}\,{d\rho_i\over dk} =
4{V_{50} \over \mpl^4}\, |T(k/k_{\rm EQ})|^2 (k\tau_0)^{-2};
\end{equation}
where $\rho_{\rm crit} =3H_0^2/8\pi G\simeq
1.05h^2\times 10^4 \eV\cmm3$ is the present value of the critical
density.  This agrees with previous results \cite{energy}
in the limit $k\ll k_{\rm EQ}$, and is shown in Fig.~3.

The spectrum of gravitational waves is cut off at wavenumber \cite{energy}
\begin{equation}
k_{\rm max}\approx {T_0\,({\cal M}^2\trh )^{1/3} \over \mpl}
\sim 10^{21}\,\left({{\cal M}\over 10^{14}\GeV}\right)^{2/3}
\left( {\trh \over 10^{14}\GeV}\right)^{1/3}\,\Mpc^{-1} ;
\end{equation}
where ${\cal M}^4$ is the vacuum energy at the very end of
inflation, $\trh$ is the reheat temperature, and $T_0=2.726\,$K
is the present temperature.  The scale $k_{\rm max}$ corresponds to the
scale that crossed outside the horizon just as inflation ended,
and is the highest frequency gravity wave produced.

If 100\% of the vacuum energy is converted into
radiation at reheating, then inflation is followed
immediately by the radiation-dominated epoch
and $\trh \simeq {\cal M}$.
For ``imperfect reheating,''
there is an epoch between the end of inflation
and the beginning of radiation domination where the energy
density of the Universe is dominated by the coherent
oscillations of the scalar field responsible for inflation,
and $\trh < {\cal M}$.
In this case, the energy density in gravitational waves
decreases as $k^{-2}$ from $k=k_*\approx T_0 \,(\trh /\mpl )$
to $k=k_{\rm max}$ \cite{inflation}; the scale $k_*$
crossed back inside the horizon just
as the Universe became radiation dominated.

The contribution of tensor perturbations to
the variance of the multipoles, which arises solely due
to the Sachs-Wolfe effect \cite{sw,cbrtensor}, is given by
\begin{equation}\label{eq:tensorlm}
\langle |a_{lm}|^2 \rangle =  36 \pi^2\,A_T \,{\Gamma (l+3)
\over \Gamma (l-1) }\, \int_0^\infty\,
|F_l(u)|^2\,|T(u/u_{\rm EQ})|^2\,{du\over u} ;
\end{equation}
where
\begin{equation}
F_l(u)  =  - \int_{(\tau_{\rm LSS}/{\tau_0})u}^u \, dy \,
\left({j_2(y )\over y }\right)\,\left[ {j_l(u-y) \over
(u -y )^2}\right] ,
\end{equation}
$u=k\tau_0$, $y=k\tau$, and $u_{\rm EQ}=k_{\rm EQ}\tau_0$.

This expression is identical to previous results
with the exception of the inclusion of the transfer
function to properly take account of
short-wavelength modes, $k \ga k_{\rm EQ}$.
Since $|T(k/k_{\rm EQ})|^2 \propto k^2$ for
$k\ga k_{\rm EQ}$, without the transfer function the
contribution from these modes has been {\it underestimated.}
In Fig.~4 we show our results for the variance
of the multipole moments, compared to those
which do not take the short-wavelength modes
into account properly.  The correction to previous
results is important even for values of $l$ as small as 10 or so.

The tensor contribution to the quadrupole CBR temperature
anisotropy is given by
\begin{equation}
\left( {\Delta T\over T_0} \right)_{Q-T}^2 \equiv
{5\langle |a_{2m}|^2\rangle\over 4\pi} \simeq 0.606 {V_{50}\over \mpl^4} ;
\end{equation}
where the integrals in the previous expressions have been
evaluated numerically; for the quadrupole term, including the transfer
function does not make a significant difference.
Using the COBE DMR measurement for the quadrupole
anisotropy (derived for scale-invariant perturbations),
$\Delta T_Q = 17\pm 4 \mu$K \cite{DMR},
as a rough upper limit to the tensor
contribution, $V_{50}$ can be bounded \cite{KW},
\begin{equation}
V_{50}\la 6.4 \times 10^{-11}\mpl^4 \simeq (3.5\times 10^{16}
\GeV )^4.
\end{equation}
This implies that inflation takes place at a subPlanck
energy scale (at least the last 50 or so e-folds that
are relevant for us) and that $\Omega_{\rm GW}$ can be at most about
$10^{-10}$.  In models of first-order inflation another,
a more potent source of gravity waves is produced by bubble
collisions during reheating, though this radiation
is peaked over a very narrow range of frequencies,
$k \approx 2\times 10^{21}\,(\trh /10^{14}\GeV )\Mpc^{-1}$ \cite{foigw}.

For reference, the Sachs-Wolfe result for the scalar modes, which
is the dominant contribution to the scalar-mode produced
anisotropy for $l\la 100$, is
\begin{eqnarray}\label{eq:scalarlm}
\langle |a_{lm}|^2\rangle & = & A\,{H_0^4\over 2\pi}\,
\int_0^\infty \,|T_S(u/u_{\rm EQ})|^2\,|j_l(u)|^2\,{du\over u} ;\nonumber \\
A & = & {1024\pi^3 \over 75H_0^{4}}\,{V_{50} \over \mpl^4 x_{50}^2};\nonumber
\\
\left( {\Delta T\over T_0} \right)_{Q-S}^2 & \equiv &
{5 \langle |a_{2m}|^2\rangle \over 4\pi } \approx {32\pi\over 45}\,
{V_{50} \over \mpl^4\,x_{50}^2};
\end{eqnarray}
where $u=k\tau_0$, $u_{\rm EQ}=k_{\rm EQ}$, and
$T_S(y)$ is the transfer function for
scalar perturbations, which depends upon the matter
content of the Universe.  For cold dark matter
the transfer function is \cite{stat}
\begin{equation}
T_S(y) = {\ln (1+0.146y)/0.146y \over
[1 + 0.242y + y^2 + (0.340y)^3 + (0.417y)^4]^{1/4}} .
\end{equation}

The scalar and tensor contributions to a given
multipole are dominated by wavenumbers $k\tau_0\approx l$.
For scale-invariant perturbations and small $l$,
both the scalar and tensor contributions to
$l^2\langle |a_{lm}|^2\rangle$, are approximately constant.
The contribution of scalar perturbations to
$l^2\langle |a_{lm}|^2\rangle$ begins
to decrease for $l=l_{\rm EQ}\sim 100$ since the scalar
contribution to these multipoles is
dominated by modes that entered the horizon before matter
domination and are suppressed by the (scalar)
transfer function.  The contribution of tensor modes to
$l^2\langle |a_{lm}|^2\rangle$ begins
to decrease for $l\sim \tau_0 /\tau_{\rm LSS} \approx \sqrt{1+\zl}$
($\sim 35$ for $\zl \sim 1100$) because the tensor contribution
to these multipoles is dominated by modes that
entered the horizon before last scattering
(and hence decayed as $R^{-1}$ until last scattering).
The behaviour of the multipole amplitudes is just
as expected from the qualitative picture, cf.~Fig.~1,
and is illustrated in Fig.~4.

\subsection{Finite thickness of the last-scattering surface}

Last scattering is not an instantaneous event that occurs
simultaneously everywhere in the Universe.  The last-scattering
surface has a finite thickness, in comoving distance
from the observer, $\sigma_x$, and in red shift, $\sigma_z$.
This fact leads to the damping of the contribution of
modes with large wavenumber because the contribution to
CBR anisotropy in a given direction averages
over last-scattering events taking place over a finite
distance, which washes out short-wavelength modes, $k\ga k_D=\sigma_x^{-1}$.
{\it This physics is not included
in our analysis.}  Since the $l$th multipole is dominated
by the contribution of wavenumbers $k\tau_0 \approx l$,
this damping only affects multipoles $l\ga \tau_0 /\sigma_x$, multipoles
that were already very small.  Thus, our neglect of the
finite thickness of the last-scattering surface does
not affect our results in an important way.

To be more specific, in the case of standard
recombination, the differential probability that the last-scattering
event occurred at comoving distance $x$---the visibility
function---can be approximated by a gaussian (see e.g.,
\cite{joneswyse})
\begin{equation}
{d{\cal P}\over dx} \equiv {ds \over dx} e^{-s (x)}
\simeq C\exp [-(x-x_{\rm LSS})^2/2\sigma_x^2],
\end{equation}
where $C\simeq 182 H_0$, $x_{\rm LSS} \simeq 2H_0^{-1}$ is
the center of the last-scattering surface,
and $\sigma_x =2H_0^{-1}/910$
is the thickness of the last-scattering surface.
(In red shift, the thickness is $\sigma_z \simeq 80$.)
Here $s (x)$ is the optical depth
from our position to a point comoving distance $x$ from us,
\begin{equation}
s (x) = \int_0^x \,n_e(x)\,\sigma_T\,{dt\over dx}\, dx,
\end{equation}
where $\sigma_T \simeq 0.66\times 10^{-24}\rcm^2$ is
the Thomson cross section and $n_e$ is the number
density of free electrons ($\approx X_e n_B$).  The damping scale associated
with the thickness of the last-scattering surface is
$k_D\tau_0 = \tau_0/\sigma_x \simeq
910$, which leads to the suppression
of the angular power spectrum for $l\ga 910$, where
it is already very small, cf. Fig.~4.

In the case of no, or only partial, recombination
the last-scattering surface is very thick, $\sigma_z
\sim \zl$.  The visibility function can be (less well)
approximated by a gaussian,
\begin{equation}
{d{\cal P}\over dx} \simeq C \exp [-(x-x_{\rm LSS})^2/2\sigma_x^2],
\end{equation}
where
$$C=1/\sqrt{2\pi}\sigma_x ;\qquad
x_{\rm LSS} \simeq 2H_0^{-1} [1-1/\sqrt{1+\zl}]; \qquad
\sigma_x \simeq 2H_0^{-1} \sqrt{12(1+\zl)}.$$
The red shift of the last-scattering surface
$$1+\zl \simeq (0.03X_e\Omega_Bh)^{-2/3}.$$
In the case of nonstandard recombination
the damping scale $k_D\tau_0 \simeq 4
\sqrt{\zl}$ is smaller---because the last-scattering
surface is thicker---but again, this damping only affects
multipoles that were very small anyway, $l\ga 4\sqrt{\zl}\gg
\sqrt{\zl}$, cf.~Fig.~5.

\subsection{Comparison with numerical results}

In Ref.~\cite{bondetal} the coupled Boltzmann equations
for the CBR intensity in
a perturbed FRW model were solved numerically to
compute the CBR anisotropy that arises due to both scalar and
tensor perturbations.  The authors of Ref.~\cite{bondetal}
have very gracious about
comparing their preliminary results with us.
Where comparison is possible, the agreement is always
qualitatively very good, though there are
quantitative disagreements:  for values of $l$ where
the multipoles are of significant size at most about
20\%.  As we shall describe, we believe that we understand the
reason for these disagreements.

For very large $l$, corresponding to $l\ga k_D\tau_0$,
the numerical results of Ref.~\cite{bondetal} fall off more
rapidly than ours; as discussed above this is because
we have not taken account of the damping due to the finite
thickness of the last scattering surface.  The discrepancy
here---though large---is of little practical concern as the
multipoles are very small for such values of $l$.

A more important discrepancy arises in cases where last-scattering
occurs before the Universe is ``very matter dominated.''  The
matter and radiation energy densities become equal at a red shift
$z_{\rm EQ} \simeq 2.4h^2\times 10^4$; for $h=0.5$ and
standard recombination $\zl$ and $z_{\rm EQ}$ differ only
by a factor of 5.4.  This means that the Universe is not
well approximated as being matter
dominated at last scattering.  In particular,
the growing mode gravity-wave perturbation is not given by
$3j_1(k\tau )/k\tau$ and
$$
\tau_{\rm LSS} = 2H_0^{-1}\left( 1+ \sqrt{R_{\rm EQ}} \right)
\left[ \sqrt{R_{\rm LSS} +R_{\rm EQ}} -\sqrt{R_{\rm EQ}}\right],
$$
which differs from the matter-dominated result, $\tau_{\rm LSS}
=2H_0^{-1}\sqrt{R_{\rm LSS}}$ by a factor of 2/3.
To correct Eq.~(\ref{eq:tensorlm}) one would have to:
(i) Use the correct expression for $\tau_{\rm LSS}$, which
is easy to do; (ii)  Modify the transfer function,
taking into account that
it is a function of red shift, which means that it can no longer
be taken out of the inner integral; and (iii) Replace
the conformal-time derivative of the growing-mode eigenfunction,
the $j_2(y)/y$ term, by the proper expression which
must be evaluated numerically and is a separate function
of $\tau$ and $k$ (points (ii) and (iii) are not
unrelated).  Needless to say, these modifications eliminate
the advantages of the analytic approach.

The parameter that controls the size of the error made by
assuming that the Universe is matter dominated at last-scattering
is $\zl /z_{\rm EQ}$.  In the worst case considered, standard recombination
and $h=0.5$, this parameter is about 0.2, which is about equal to the maximum
difference between our analytical results and
the numerical results of Ref.~\cite{bondetal}.
When $\zl /z_{\rm EQ}$ is smaller, e.g., larger
Hubble constant or nonstandard recombination,
the discrepancy is much smaller.  For example, with $h=0.5$
and $\zl = 76$, the differences between the numerical results
of Ref.~\cite{bondetal} and our analytical
results is only a few per cent (until the multipoles
become very small due to the damping associated with
the finite thickness of the last-scattering surface).
The advantage
of the analytic approach is ease of calculation, particularly
the ability to study a variety of scenarios.

\subsection{Generalized for tilt}

We now extend our results to the general case where
the inflationary perturbations are not precisely scale
invariant.  Since the ratio of the tensor to scalar
contribution to the CBR quadrupole anisotropy increases
with tilt, $\langle |a_{2m}^T|^2\rangle /
\langle |a_{2m}^S|^2\rangle \approx -7n_T$ ($n_T=0$ for scale invariance),
the case of non scale-invariant
perturbations is very relevant in discussing tensor
perturbations.  The deviations from scale invariance
are most significant for steep potentials (e.g., exponential potentials)
and potentials whose steepness changes rapidly (e.g., low-order
polynomial potentials).
To lowest order in the deviation from scale invariance
everything can be expressed
in terms of the value of the potential $V_{50}$,
its steepness $x_{50} =(\mpl V^\prime /V)_{50}$,
and the change in its steepness $x_{50}^\prime$,
all evaluated about 50 e-folds before the end of inflation
where the scales of astrophysical interest crossed
outside the horizon during inflation \cite{turner,ls}.

Beginning with the tensor perturbations and using our
previous notation, the power spectrum is given by
\begin{eqnarray}
P_T(k) & = & A_T k^{n_T -3} |T(k/k_{\rm EQ})|^2
\left[ {3j_1(k\tau )\over k\tau}\right]^2 ; \nonumber \\
A_T & = & {8 \over 3\pi }\, {V_{50} \over \mpl^4}\,
{[\Gamma ({3\over 2} - {1\over 2}n_T)]^2 \over
(1- {5\over 6}n_T) 2^{n_T}
[\Gamma ({3\over 2})]^2}\, k_{50}^{-n_T} ;  \nonumber \\
n_T & = &  -{x_{50}^2\over 8\pi} .
\end{eqnarray}
Here $k_{50}$ is the scale that crossed outside the
horizon 50 e-folds before the end of inflation,
$n_T$ measures the deviation from scale invariance,
and the expression for $A_T$ includes the ${\cal O}(n_T)$ correction.
(All formulas simplify if the potential and its derivatives
are evaluated at the point where the present horizon
scale crossed outside the horizon, i.e., $k_{50}\tau_0 \approx 1$.)
The variance of the multipole amplitudes is given by
\begin{equation}\label{eq:tensor1lm}
\langle |a_{lm}|^2 \rangle =  36 \pi^2 \,A_T\, {\Gamma (l+3)
\over \Gamma (l-1) }\,\tau_0^{-n_T}\, \int_0^\infty\,
u^{n_T} \, |T(u/u_{\rm EQ})|^2\,|F_l(u)|^2\,{du\over u} ;
\end{equation}
where, as before,
\begin{equation}
F_l(u)  =  - \int_{(\tau_{\rm LSS}/\tau_0)u}^{u} \, dy \,
\left({j_2(y )\over y }\right)\,\left[ {j_l(u-y) \over
(u-y)^2}\right] ;
\end{equation}
$u=k\tau_0$, $y=k\tau$, and $u_{\rm EQ}=k_{\rm EQ}\tau_0$.

In Fig.~6 we show the scalar and tensor contributions to
the angular power spectrum for spectra that are tilted
by the same amount, $n-1 = n_T
=-0.15$.  This is an interesting case because
the scalar and tensor contributions
to the quadrupole anisotropy are essentially equal
\cite{tilt}.  The effect of
scale noninvariance is to tilt the angular power spectra,
by approximately a factor of $l^{n-1}$ for scalar and $l^{n_T}$ for tensor.

The analogous expressions for scalar perturbations are
\begin{eqnarray}\label{eq:scalar1lm}
A & = & {1024\pi^3 \over 75H_0^{3+n}}\,{V_{50} \over \mpl^4 x_{50}^2}
\,{\left\{ \Gamma [{3\over 2} -{1\over 2}(n-1)]\right\}^2
\over  [1+{7\over 6}n_T - {1\over 3}(n-1)]
2^{n-1} [\Gamma ({3\over 2})]^2}\,k_{50}^{1-n} ;\nonumber \\
n & = &  1 -{x_{50}^2\over 8\pi} + {\mpl x_{50}^\prime\over 4\pi} ;\nonumber\\
\langle |a_{lm}|^2\rangle  & = & A\,{H_0^4\over 2\pi}\,\tau_0^{1-n}\,
\int_0^\infty \,u^{n-1}\,|T_S(u/u_{\rm EQ})|^2\,|j_l(u)|^2\,{du\over u};
\end{eqnarray}
where $u=k\tau_0$, $u_{\rm EQ}=k_{\rm EQ}$,
and $T_S(u/u_{\rm EQ})$ is the transfer function for scalar perturbations.

By numerically integrating Eqs.~(\ref{eq:tensor1lm},\ref{eq:scalar1lm})
we can obtain expressions for the quadrupole anisotropy due
to scalar and tensor perturbations in the non scale-invariant case:
\begin{eqnarray}
{5\langle |a_{lm}^S|^2\rangle \over 4\pi }   & = &
2.22 {V_{50}\over \mpl^4 x_{50}^2} \left( 1+1.1(n-1) +{7\over 6}
[n_T -(n-1)] \right)     ; \nonumber \\
{5\langle |a_{lm}^T|^2\rangle \over 4\pi}    & = &
0.606 {V_{50}\over \mpl^4 } \left( 1+1.2n_T \right) ;\nonumber \\
{\langle |a_{2m}^T|^2 \rangle \over \langle |a_{2m}^S|^2
\rangle} & \simeq & -7n_T \left( 1 +1.1n_T+ 0.1[n_T-(n-1)]\right) .
\end{eqnarray}
We have taken $k_{50}\tau_0 =1$, $\zl =1100$ (the results
change very little for $\zl =76$), and for the
scalar case $h=0.5$.  Using these
expressions, the fact that the scalar and tensor contributions
to the quadrupole anisotropy add incoherently,
and the COBE DMR quadrupole
measurement, we can solve for the
variance of the tensor quadrupole, equivalently $V_{50}
/\mpl^4$, in terms of the tensor tilt $n_T$:
\begin{eqnarray}\label{eq:tensornorm}
\langle|a_{2m}^T|^2\rangle & \simeq & {9.8\times 10^{-11}\over
1 - 0.14n_T^{-1}} ;\nonumber \\
{V_{50}\over \mpl^4} & \simeq & {6.4\times 10^{-11} \over
1-0.14 n_T^{-1}}.
\end{eqnarray}
These expressions indicate that the more tilted the
gravity-wave spectrum is, the larger its amplitude
is, as noted earlier \cite{tilt}.

The energy density in the stochastic background of inflation-produced
gravitational waves today is given by
\begin{eqnarray}\label{eq:gwtilt}
{d\Omega_{\rm GW} \over d\ln k} & = & 4{V_{50} \over \mpl^4}\,
{[\Gamma ({3\over 2} - {1\over 2}n_T)]^2 \over (1-{5\over 6}n_T)
2^{n_T} [\Gamma ({3\over 2})]^2}\, (k/k_{50})^{n_T}
|T(k/k_{\rm EQ})|^2\,(k\tau_0)^{-2} ;\nonumber \\
& \simeq & {2.6\times 10^{-10} \over 1-0.14n_T^{-1}}\,
(1 + 0.1n_T) \, (k/k_{50})^{n_T}
|T(k/k_{\rm EQ})|^2\,(k\tau_0)^{-2} ;
\end{eqnarray}
where the second expression follows from using the COBE DMR
normalization to express the energy density in gravity waves in
terms of the tilt parameter $n_T$ alone.
In Fig.~3 we show the spectrum of stochastic gravitational waves
for $n_T=-0.003, -0.03, -0.3$, using the COBE quadrupole normalization.

The total energy density in gravity waves increases with tilt (i.e.,
$n_T<0$), as does the tensor contribution to the CBR quadrupole
anisotropy.  However, this is not the entire story; the most
sensitive ``direct'' probes of gravity waves,
millisecond pulsars \cite{MSP} and future Laser Interferometer
Gravity-wave Observatories (LIGOs)
\cite{LIGO}, are only sensitive to gravity waves with
very large wavenumber,
$k\tau_0 \equiv e^N$ with $N\sim 26$ for millisecond
pulsars and $N\sim 48$ for envisaged LIGO detectors.
Because of the $(k/k_{50})^{n_T}$ factor in Eq.~(\ref{eq:gwtilt})
tilt depresses the energy density in gravity
waves on the relevant scales.  To
be more specific, for $k\gg k_{\rm EQ}$ and $h=0.5$
\begin{equation}
{d\Omega_{\rm GW}(k=e^N\tau_0^{-1})
\over d\ln k} \simeq 1.9\times 10^{-14}
{n_T\, e^{n_TN} \over n_T -0.14} .
\end{equation}
It is simple to show that the energy density in gravity waves
on the scale $k = e^N\tau_0^{-1}$ is {\it maximized} for a value
of $n_T\approx -1/N$, at a value of about $5\times 10^{-14}/N$.
While the amount of tilt that maximizes the energy density
in gravity waves on the scales relevant to both millisecond
pulsars, $n_T\simeq -0.04$, and LIGOs, $n_T\simeq -0.02$, is
quite reasonable in the context of inflationary models
\cite{turner}, the predicted energy density in gravity waves is well
below the sensitivity of either detector, about
$\Omega_{\rm GW}\sim 10^{-10}$ for advanced
LIGO detectors and $\sim 10^{-7}$ currently
for millisecond pulsars.  Thus, direct
detection of the stochastic background of gravitational
waves does not seem promising in the foreseeable future.

\section{Discussion}

The tensor contribution to the variance
of the multipole amplitudes depends significantly
upon the red shift of the last-scattering surface and less importantly
upon the red shift of matter-radiation equality.
For scale-invariant gravity-wave perturbations $l^2\langle |a_{lm}|^2\rangle$
is roughly constant for $l\la \sqrt{1+\zl}$;
then decreases as $l^{-4}$ for $l\la \sqrt{1+z_{\rm EQ}}$;
and for $l\ga \sqrt{1+z_{\rm EQ}}$ decreases as $l^{-2}$.
In the case of non scale-invariant tensor perturbations
these results are modified by a factor of $l^{n_T}$.

In Fig.~5 we show the tensor contribution
to the angular power spectrum for no recombination and
$\zl =76$.  The dramatic fall off occurs
at a relatively small value of $l$, around 10.  Thus,
the tensor angular power spectrum can, in principle,
be used to discriminate between standard recombination
and no recombination, though due to cosmic variance,
the finite ``multipole resolution'' of experiments, and
the difficulty of separating the scalar and tensor
contributions to CBR anisotropy
this is by no means a simple task.  The angular power spectrum also
depends upon the deviation of the tensor perturbations
from scale invariance, both in its amplitude relative
to the scalar perturbations and its dependence upon
$l$.  The angular power spectrum for tilted tensor
perturbations is shown in Fig.~6.

Finally, consider the cold dark matter + cosmological
constant model ($\Lambda$CDM), proposed to reconcile
a number of discrepancies of the CDM model with observational
data \cite{lcdm}.  It is characterized by:
$\Omega_B \simeq 0.05$, $\Omega_{\rm cold} \simeq 0.15$,
$\Omega_0=\Omega_B+\Omega_{\rm cold} =0.2$,
$\Omega_\Lambda \simeq 0.8$, and $h\simeq 0.8$.
For tensor perturbations most of the Sachs-Wolfe
integral for the anisotropy arises near the last
scattering surface where the effect of a cosmological
constant is negligible.  (This is not the case for
scalar perturbations, and the formula for the
Sachs-Wolfe contribution must be modified significantly
\cite{lambdacbr}.)  Thus
Eq.~(\ref{eq:tensorlm}) for the contribution of
tensor perturbations to the
angular power spectrum is unchanged, with the following substitutions:
\begin{eqnarray}
k_{\rm EQ} & = & {H_0 \sqrt{\Omega_0/ R_{\rm EQ} }\over
2\sqrt{2}-2} \simeq 30 H_0 ; \nonumber \\
R_{\rm EQ} & = & 4.18\times 10^{-5}\,(\Omega_0 h^2)^{-1}
\simeq 3.27\times 10^{-4}; \nonumber \\
\tau_0 & \simeq & H_0^{-1}\int_0^1 {dR \over \sqrt{\Omega_0 R
+ \Omega_0 R_{\rm EQ} +\Omega_\Lambda R^4}} \simeq 3.89 H_0^{-1}; \nonumber \\
\tau_{\rm LSS} & \simeq & 2\Omega_0^{-1/2}H_0^{-1}/\sqrt{1+\zl}
\simeq 0.135 H_0^{-1}.
\end{eqnarray}

The angular power spectrum is shown in Fig.~7 for
standard recombination in the $\Lambda$CDM model.
Unfortunately, the difference between the $\Lambda$CDM and CDM models
is not great.

Finally, it is also straightforward to modify
our results for the energy density in gravitational waves
today for the $\Lambda$CDM model:
\begin{equation}
{d\Omega_{\rm GW} \over d\ln k} = 4\Omega_0^2 {V_{50} \over \mpl^4}\,
{[\Gamma ({3\over 2} - {1\over 2}n_T)]^2 \over (1-{5\over 6}n_T)
2^{n_T} [\Gamma ({3\over 2})]^2}\, (k/k_{50})^{n_T}
|T(k/k_{\rm EQ})|^2\,(k/2H_0^{-1})^{-2} .
\end{equation}
Since the main change is to {\it reduce} the energy density
on a given scale by a factor of $\Omega_0^2\sim 0.04$, our
conclusions about the direct detection of inflation-produced
gravity waves remains.

\vskip 1.5cm

\noindent
We thank Paul J.~Steinhardt for helping us make detailed
comparisons between our work and that in
Ref.~\cite{bondetal}.  This work
was supported in part by the DOE (at Chicago and Fermilab) and by
the NASA through NAGW-2381 (at Fermilab).

\newpage
\centerline{\large\bf FIGURE CAPTIONS}
\bigskip\bigskip

\noindent{\bf Figure 1:}  The qualitative behaviour expected
for CBR anisotropy arising from scalar and tensor perturbations.
The horizon-crossing amplitudes of the scale and tensor
perturbations are taken to be
$\varepsilon_S \propto \lambda^{(1-n)/2}$
and $\varepsilon_T\propto \lambda^{-n_T/2}$ respectively
(scale invariance corresponds to $n-1 = n_T = 0$).

\bigskip
\noindent{\bf Figure 2:} The evolution of the transfer
function for gravity-wave perturbations; from top to
bottom, the transfer function computed at red shifts
$z=0,30,100,300,1000$.

\bigskip
\noindent{\bf Figure 3:}  The energy density of the stochastic
background of inflation-produced gravitational waves, in the scale-invariant
limit (broken curve, arbitrary normalization),
and for $n_T=-0.003, -0.03, -0.3$, normalized to the COBE DMR quadrupole
(solid curves).

\bigskip
\noindent{\bf Figure 4:}  The tensor contribution to the
angular power spectrum normalized to the quadrupole,
and for reference the scalar contribution normalized to
the quadrupole (broken curve).  Tensor results are shown with (lower
curve) and without (upper curve) the transfer function;
in all cases $\zl =1100$ and $h=0.5$.  (a)  For $l\le 100$;
dotted curve shows tensor results with transfer function for
$h=1.0$.  (b)  For $l\le 1000$.

\bigskip
\noindent{\bf Figure 5:}  The tensor contribution to the
angular power spectrum normalized to the quadrupole,
for $\zl = 76$ (solid curve) and for comparison $\zl =1100$
(broken curve); in both cases $h=0.5$.

\bigskip
\noindent{\bf Figure 6:} Same as Fig.~4, except for non-scale
invariant perturbations, $n-1 = n_T = -0.15$.  The results
here correspond roughly to those in Fig.~4 ``tilted'' by a factor
of $(l/2)^{-0.15}$.

\bigskip
\noindent{\bf Figure 7:}  The tensor contribution to the
angular power spectrum normalized to the quadrupole,
for cold dark matter + cosmological constant with
$\zl =1100$, $\Omega_\Lambda =0.8$, and $h=0.8$ (solid curve)
and for comparison, $\zl=1100$, $\Omega_0$, and $h=0.5$
(broken curve).


\vskip 2cm


\begin{thebibliography}  {inflation}

\bibitem{scalar}  A.H.~Guth and S.-Y.~Pi, {\it Phys. Rev. Lett.}
{\bf 49}, 1110 (1982); A.A.~Starobinskii, {\it Phys. Lett. B}
{\bf 117}, 175 (1982); S.W.~Hawking, {\it ibid} {\bf 115}, 295 (1982);
J.M.~Bardeen, P.J.~Steinhardt, and M.S.~Turner, {\it Phys. Rev. D}
{\bf 28}, 679 (1983).

\bibitem{tensor} V.A.~Rubakov, M.~Sazhin, and A.~Veryaskin,
{\it Phys. Lett. B} {\bf 115}, 189 (1982); R.~Fabbri and
M.~Pollock, {\it ibid} {\bf 125}, 445 (1983); L.~Abbott
and M.~Wise, {\it Nucl. Phys. B} {\bf 244}, 541 (1984);
B.~Allen, {\it Phys. Rev. D} {\bf 37}, 2078 (1988); L.P.~Grishchuk,
{\it Phys. Rev. Lett.} {\bf 70}, 2371
(1993) and earlier references therein.

\bibitem{inflation} For a textbook discussion of inflation
see e.g., E.W.~Kolb and M.S.~Turner, {\it The Early
Universe} (Addison-Wesley, Redwood City, CA, 1990), Ch.~8.

\bibitem{tilt} R.~Davis et al., {\it Phys. Rev. Lett.}
{\bf 69}, 1856 (1992); F.~Lucchin, S.~Matarrese, and
S.~Mollerach, {\it Astrophys. J.} {\bf 401}, L49 (1992); D.~Salopek,
{\it Phys. Rev. Lett.} {\bf 69}, 3602 (1992); A.~Liddle and D.~Lyth,
{\it Phys. Lett. B} {\bf 291}, 391 (1992);
T.~Souradeep and V.~Sahni, {\it Mod. Phys. Lett. A} {\bf 7}, 3541 (1992).

\bibitem{turner} M.S.~Turner, astro-ph/9302013 (FERMILAB-Pub-93/026-A).

\bibitem{reconstruct} Attempts at the
reconstruction of the inflationary potential
from observational data include, H.M.~Hodges and
G.R.~Blumenthal, {\it Phys. Rev. D} {\bf 42}, 3329 (1990);
and more recently,
E.~Copeland, E.W.~Kolb, A.~Liddle, and J.~Lidsey,
FERMILAB-Pub-93/029-A (1993).

\bibitem{sw} R.K.~Sachs and A.M.~Wolfe, {\it Astrophys. J.}
{\bf 147}, 73 (1967).

\bibitem{cbranisotropy} See e.g., G.~Efstathiou,
in {\it The Physics of the Early Universe}, eds.~J.A.~Peacock,
A.F.~Heavens, and A.T.~Davies (Adam Higler, Bristol, 1990);
J.R.~Bond and G.~Efstathiou, {\it Mon. Not. R.
astron. Soc.} {\bf 226}, 655 (1987); J.R.~Bond et al.,
{\it Phys. Rev. Lett.} {\bf 66}, 2179 (1991); P.J.E.~Peebles,
{\it Large-scale Structure of the Universe} (Princeton University
Press, Princeton, 1980).

\bibitem{cbrtensor} A.A.~Starobinskii, {\it JETP Lett.}
{\bf 11}, 133 (1985);  L.~Abbott and M.~Wise, {\it Nucl. Phys. B}
{\bf 244}, 541 (1984); M.~White, {\it Phys. Rev. D} {\bf 46}, 4198 (1992);
R.~Fabbri, F.~Lucchin, and S.~Matarrese, {\it Phys. Lett. B}
{\bf 166}, 49 (1986).

\bibitem{bondetal} R.~Crittenden et al., astro-ph/9303014;
P.J.~Steinhardt, private communication (1993).

\bibitem{joneswyse}  B.~Jones and R.~Wyse, {\it Astron.
Astrophys.} {\bf 149}, 144 (1985).

\bibitem{walkeretal} T.P.~Walker et al., {\it Astrophys. J.}
{\bf 376}, 51 (1991).

\bibitem{silk} N.~Tegmark and J.~Silk, {\it Astrophys. J.},
in press (1993).

\bibitem{energy} V.A.~Rubakov, M.V.~Sazhin, and A.V.~Veryaskin,
{\it Phys. Lett. B} {\bf 115}, 189 (1982);
R.~Fabbri and M.~Pollock, {\it ibid} {\bf  125}, 445 (1983);
B.~Allen, {\it Phys. Rev. D} {\bf 37}, 2078
(1988); V.~Sahni, {\it ibid} {\bf 42}, 453 (1990);
M.~White, {\it ibid} {\bf 46}, 4198 (1992).

\bibitem{DMR} G.~Smoot et al., {\it Astrophys. J.} {\bf 396}, L1 (1992);
E.L.~Wright, {\it ibid} {\bf 396}, L3 (1992).

\bibitem{KW}  L.~Krauss and M.~White, {\it Phys. Rev. Lett.}
{\bf 69}, 869 (1992).

\bibitem{foigw}  M.S.~Turner and F.~Wilczek, {\it Phys.
Rev. Lett.} {\bf 65}, 3080 (1990);
A.~Kosowsky, Michael S.~Turner, and R.~Watkins
{\it ibid} {\bf 69}, 2026 (1992).

\bibitem{stat} J.M.~Bardeen et al., {\it Astrophys. J.}
{\bf 304}, 15 (1986).

\bibitem{ls} D.H.~Lyth and E.D.~Stewart, {\it Phys. Lett. B}
{\bf 274}, 168 (1992); E.D.~Stewart and D.H.~Lyth, {\it ibid},
in press (1993).

\bibitem{MSP} D.R.~Stinebring, M.F.~Ryba, J.H.~Taylor,
and R.W.~Romani, {\it Phys. Rev. Lett.} {\bf 65}, 285 (1990).

\bibitem{LIGO} A.~Abramovici et al., {\it Science} {\bf 256},
325 (1992); J. E. Faller et al., {\it Adv. Space Res.} {\bf 9},
107 (1989).

\bibitem{lcdm}  M.S.~Turner, G.~Steigman, and L.~Krauss,
{\it Phys. Rev. Lett.} {\bf 52}, 2090 (1984); M.S.~Turner,
{\it Physica Scripta} {\bf T36}, 167 (1991); P.J.E.~Peebles,
{\it Astrophys. J.} {\bf 284}, 439 (1984); G.~Efstathiou et al.,
{\it Nature} {\bf 348}, 705 (1990).

\bibitem{lambdacbr} K.~Gorski, J.~Silk, and N.~Vittorio,
{\it Phys. Rev. Lett.} {\bf 68}, 733 (1992).

\end{thebibliography}
\end{document}